# WEARDA: Recording Wearable Sensor Data for Human Activity Monitoring

RICHARD M.K. van DIJK
DANIELA GAWEHNS
MATTHIJS VAN LEEUWEN

*Author affiliations can be found in the back matter of this article

## ABSTRACT

We present WEARDA,[1] the open source WEARable sensor Data Acquisition software package. WEARDA facilitates the acquisition of human activity data with smartwatches and is primarily aimed at researchers who require transparency, full control, and access to raw sensor data. It provides functionality to simultaneously record raw data from four sensors—tri-axis accelerometer, tri-axis gyroscope, barometer, and GPS—which should enable researchers to, for example, estimate energy expenditure and mine movement trajectories. A Samsung smartwatch running the Tizen OS was chosen because of 1) the required functionalities of the smartwatch software API, 2) the availability of software development tools and accessible documentation, 3) having the required sensors, and 4) the requirements on case design for acceptance by the target user group. WEARDA addresses five practical challenges concerning preparation, measurement, logistics, privacy preservation, and reproducibility to ensure efficient and errorless data collection. The software package was initially created for the project "Dementia back at the heart of the community",[2] and has been successfully used in that context.

# (1) OVERVIEW

## 1.1 INTRODUCTION

In the social sciences (e.g., [6, 5]) and health sciences (e.g., [3, 8]), consumer smartwatches are often used to measure human activity levels or movement. These smartwatches are typically chosen because of their convenience: they are affordable, small, and have a plethora of sensors built-in. The collected wearable sensor data can be used, for example, to design better urban infrastructure [7], to monitor patients [1], or to assess the effect of lifestyle behavior on long-term health outcomes [4].

Recording raw sensor data from smartwatches often poses a problem to researchers though, especially when working with consumer-grade devices. While they are usually cheaper than research devices, consumer-grade devices often offer a more closed environment and record data that is already aggregated or otherwise processed (e.g., by applying some unknown form of activity recognition). Depending on the brand, researchers can opt for paying external providers to aggregate and download data from them or to use a limited set of data points that is made accessible to consumers [3]. In practice, however, researchers usually require complete access to the raw data while cloud services are not an option due to privacy reasons.

To address these needs, we present WEARDA, the open-source WEARable sensor Data Acquisition software package that is primarily aimed at researchers who require transparency, full control, and access to raw sensor data. By running as an app on a smartwatch, WEARDA allows the recording of raw sensor data from a consumer smartwatch bypassing cloud-based or third-party privacy-sensitive services. The software can be used with Samsung smartwatches running the Tizen operating system[3] and can record and access raw data from the tri-axis accelerometer, tri-axis gyroscope, barometer, and GPS sensors. While WEARDA was originally developed for real-life data collection, it can also be used to collect data in controlled settings. We chose a Samsung smartwatch[4] running the Tizen OS because of 1) the required functionalities of the smartwatch software API, e.g., the possibility to read the raw sensor data and write this to local storage; 2) the availability of embedded software development tools and accessible documentation; 3) having the required sensors for human activity monitoring and localization, and 4) requirements on case design for acceptance by the target user group of the project the software was initially developed for, i.e., people with dementia.

Collecting activity data on human subjects—either in real life or in a controlled laboratory environment—is a difficult and error-prone task. Without a thought-out process supported by the software, problems can occur during all three main phases of the data collection, i.e., preparation, the actual measurement, and data logistics. Besides the practical challenges in these three phases, privacy preservation and reproducibility are two additional areas of attention. Table 1 gives an overview of how WEARDA addresses these five practical challenges; this is described in more detail in Section 2.1.

## 1.2 DATA COLLECTION USING WEARDA

By addressing the practical challenges as shown in Table 1, WEARDA aims to make it as simple as possible to smoothly prepare for data collection, then robustly collect sensor data while preserving privacy, and finally upload the data to an on-site laptop. The workflow of using the software can be summarised as follows.

The first step is to create a configuration file with the desired settings, upload this to all watches that will be used in the study, and check and test all watches (*preparation* phase). Once on site, the researcher activates each watch using the WEARDA sensor app: remove all previously collected data, enter a person-id, start the measurement, and then let the subject wear the watch on their wrist (*measurement* phase). Once the data collection is complete, a laptop is used to copy the data from each watch to a folder, first sorted by person-id, and after that by date time (*data logistics* phase).

The following section describes the implementation and architecture of the WEARDA software. It is followed by a section regarding the quality control of the package, where several tests and techniques are provided to ensure the high quality of the delivered software. In the availability section, details on the WEARDA software system requirements and dependencies are given, as well as information on how to obtain the software. The last section describes the software reuse potential and provides an overview of how to configure and run WEARDA on a Samsung watch.

# 2 IMPLEMENTATION AND ARCHITECTURE

The WEARDA software package consists of two main components: 1) the interactive "Sensor application" front-end that can be operated through the touchscreen of the watch, and 2) a data collection "Sensor service" that runs in the background on the watch. These components implement five practical challenges described first in the next subsection.

## 2.1 THE FIVE PRACTICAL CHALLENGES

The first practical challenge is the efficient and errorless preparation and configuration of smartwatches. We, therefore, chose a configuration file that can be prepared beforehand and can easily be uploaded onto the watches while they are charged before the measurements take place.

van Dijk et al. *Journal of Open Research Software* DOI: 10.5334/jors.454    3| CHALLENGES | PREPARATION | MEASUREMENT | LOGISTICS | PRIVACY | REPRODUCIBILITY |
|---|---|---|---|---|---|
| Configuration file 1 | v | | | | v |
| Watch identifier 2 | v | | v | v | v |
| Automatic correction of configuration file 3 | | v | | | v |
| Removal of files by UI watch 4 | v | v | | v | |
| Not easy to remove file by UI watch 5 | v | v | | | |
| Person identifier 6 | | v | v | v | |
| Storage of all sensor data 7 | | v | | | |
| Easy to enter person identifier 8 | | v | | | |
| Data storage for multiple days 9 | | v | v | | |
| Hard to inadvertently stop data collection 10 | | v | | | |
| Use of stand-alone laptop 11 | | v | v | v | |
| Quick upload of files to laptop 12 | | v | v | | |
| Wireless upload to laptop 13 | | | v | v | |
| Person id, date, time, and watch id in data filename 14 | | | v | | v |
| Stop measurement only with shutdown 15 | | v | | | |
| Metadata in first line of files 16 | | | | | v |
| Copy of used configuration file 17 | | | | | v |
| Label private data 18 | | | | v | |
| Change configuration file while measuring 19 | | v | | | |

**Table 1** Practical challenges and how WEARDA addresses them.

The second practical challenge is efficient and errorless measurement execution. The WEARDA software collects data in the background, invisible to the subject, and it cannot be stopped to prevent unintended interventions by the subject or researcher. The stored measurements on the watch can be copied to a standalone laptop wireless, and later removed, with Tizen software running on this laptop. We chose the digital spinner wheel widget that shows only possible id's while swiping (errorless entry) as opposed to a free format data entry.

The third practical challenge of data collection is efficient and errorless logistics. The measurements are linked to a unique identifier of the watch and are accompanied by the (corrected) configuration settings used by that particular watch. The WEARDA software replaces invalid configuration settings by default values: it will not block the start of the measurements, nor produce error messages, if errors are detected in the original configuration settings. With this correction mechanism, the researcher will not be confronted with the absence of measurements later, after intensive data collection efforts on site. We chose this correction error handling because giving notice of warning or errors could confuse subjects and researchers.

The fourth practical challenge is the preservation of privacy. We used the GPS geo-fencing API of Tizen OS by configuring a privacy circle as geofence around the area where the data collection takes place. All data collected outside of this circle are marked in the data file, but still recorded and can easily be removed in a post-processing step. For our use case, data integrity out-weight privacy concerns and we opted for gathering all data. Other use cases might require data to not only be marked but deleted before downloading. This would require additional development of the software.

The fifth practical challenge is to take measures for the reproducibility of the data collection. The configuration file does not only ease the process of setting up watches for data collection; it also helps to make the data collection reproducible for others by making data collection choices transparent and traceable.

The technical descriptions of the 'Sensor application' and 'Sensor service' components, in the next two subsections, indicate which software functions are included to address the practical challenges. This is indicated between brackets, as [x,P] – preparation, [x,M] – measurement, [x,L] – logistics, [x,Pr] – privacy, and [x,R] – reproducibility, where x represents the software function listed in Table 1.

## 2.2 SENSOR APPLICATION

The software package contains the Tizen OS widget application "liacs.sensorapplication", visible in the list of applications with the title "Sensors" and a digital brain as the icon shown in Figure 1.

### Application front-end

The application carries a spinner entry widget control to enter the person-id, a number between 000 and 999 [6,8,M], and two buttons: the RESTART button to start or restart measurements [M], and the CLEAN button to remove all measurements from the watch [4,P,M,Pr]. The CLEAN button can be activated by pressing it three times to prevent unintended removal when accidentally pressing the button [5,P,M]. The application can be closed by pressing the back button on the side of the watch. This will not influence the measurements and allows the participant to use other applications on the watch. We did not supply a STOP button because of the fear that participants would press this button unintended [10,M]. Thus, the measurements can only be stopped by a full shutting down of the watch. After several measurements, the battery will be depleted, and the watch shuts down automatically. While re-charging, the data can be downloaded, removed, and the watch is ready for the next participant. Shutting the watch down reduces power consumption, and thus, the charge duration will be shortened for the next experiment to start [15,M]. In a data collection setting where the data needs annotation, the person-id can be used as an annotation label. For example, if different types of daily activities need labeling, such as washing dishes, cutting vegetables, or biking, the person-id could be used (e.g., ID 001 = washing dishes, 002 = cutting vegetables, et cetera).

### Measurement files

The measurement records are stored in comma-separated values sensor files, and the file name contains the person-id, date, time, watch-id, and sensor types [14,L,R]. This information also appears in the first line of these files [16,R]. The researcher retrieves the files by use of the Smart Development Bridge Tool (SDB): a command line tool that is part of the Tizen Studio package. It is advised to retrieve the files daily while collecting data, for example, to validate the data in the evening, to prevent loss of data because watches can get lost or damaged. In principle, the watch can store 0.5 GB of data, which should be enough for 7 days of 12 hours of measurements a day. So, it could be possible to monitor real-life subjects during the day for one week continuously without the intervention of the researcher to download the data frequently during this period [9,M]. However, the battery will be depleted after 12 hours, so the subject should recharge the watch battery daily in this data collection scenario.

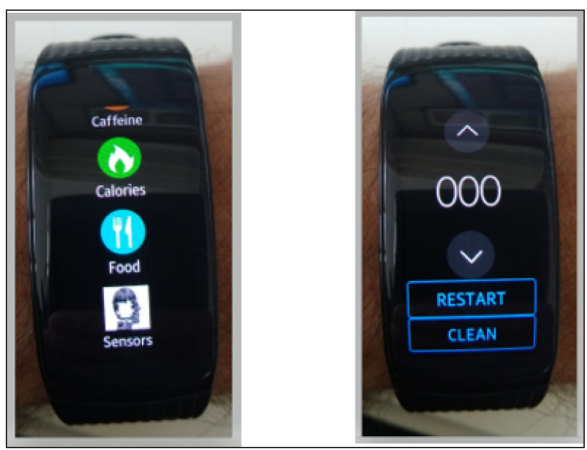

**Figure 1** Left: the icon of the sensor application in the Tizen OS menu, right the sensor application with the person-id spinner, restart, and clean button.

### Files separation

Sensor values can be recorded with different frequencies. The motion sensors are usually read out with high frequency (25–50 Hz), while the barometer, battery, and GPS (Global Positioning System) have lower frequencies (0.25 – 2 Hz). For every frequency, a separate file is created. In that way, there is no need for sensors with low-frequency rates to be recorded many times as duplicates, if these recordings are mixed with sensors that are recorded with higher frequency rates. The separation of files thus reduces power consumption and memory usage, because it reduces also the number and size of the recordings [9,M,L]. Sensor values are stored in the comma-separated value file format.

### Central clock

The application has one central clock. The time recordings in several sensor files can be aligned with this central time reference. It is advised to synchronize the time of the watches with internet-available clocks before the data collection. Precise time synchronization can be established automatically by connection to the internet via Bluetooth or WIFI.

### Configuration file

The measurements are configured with a configuration file that is uploaded to the watch and placed in an application-independent, accessible file system directory [1,M,R]. Its contents are read before starting the measurements. The used configuration settings and the software package version is saved in a metafile that has an identical structure to the original configuration file. This file becomes an integral part of the sensor files generated by each series of measurements [17,R]. Errors in the configuration file are corrected to default settings instead of giving a warning [3,M,R]. A configuration file is uploaded to the watches with the SDB command line tool of the Tizen Studio package.

### Privacy preservation

The sensor service writes sensor values to the sensor files continuously. This is independent of the privacy preservation settings. However, if the privacy preservation settings are activated, the sensor values are enriched with a privacy label indicating that the watch was inside the public area ('I'), or outside the public area ('P') [18,Pr]. In case the GPS coordinates are not available, the privacy preservation data can not be collected and the privacy label is '?'. After data collection, the collected files can be processed offline to preserve the privacy of the real-life subject by removing the data records which indicate the label outside the public area 'P'. The reason to store collected data on the watch – while the subject is outside the public area in the first place – and remove privacy-sensitive data later with a post-processing step is 1) it could be that the GPS geofence is not accurate enough or 2) the GPS base point is not set correctly which could result in empty data collections if the data was not stored at all. So, this step serves the wish for overcoming human errors [7,M].

The public area can be set by the configuration file by setting a circle on the map by defining a GPS middle point expressed in longitude and latitude coordinates, and its radius in meters. Privacy is only preserved if the GPS sensor is switched on and the sensor can establish the location of the watch, which is more likely when the watch is worn outdoors. We intend to extend WEARDA with a function to remove the privacy data during the upload to a standalone laptop based on the active configuration file.

### Non-blocking error handling

Finally, we mention the non-blocking error-handling approach. Errors are corrected if possible, or at least will be bypassed to prevent sudden unexpected stops. For example, if the configuration file contains values being out of range, they will be replaced by typical or default values [3,M,R]. For the release version of the WEARDA package, errors are not shown to the user nor they are logged. Logging in general (not only for errors) is switched off to reduce power consumption and memory use. Only, the debug version of the WEARDA package will log information about the program flow, or warnings, or errors.

## 2.3 SENSOR SERVICE

The software package contains a Tizen OS widget service, "liacs.sensorservice", which runs in the background, neither visible in the list of applications nor visible via the display of the watch. This design makes continuous measurements possible since Tizen applications become idle after a brief time to reduce the power consumption of the display, while services only become idle if the battery is (almost) empty [10,M].

### Processing messages

The sensor service processes the messages RESTART and CLEAN sent by the sensor application. After switching on the smartwatch, the sensor service is activated and waiting for these messages. After one valid message is received, the service starts the recording of the sensor values and stores it in comma-separated values files. The measurement is based on the contents of the uploaded configuration file. Each RESTART message will reload the configuration file found in its upload folder. This allows the researcher to change the configuration of watches while the measurement is still running [19,M].

Figure 2 shows a state diagram with transitions with the format "event – actions", as a formal description of the states, and relevant state events between the watch application manager, the sensor application, and the service.

| FIELD NAME | FIELD TYPE | POSSIBLE VALUES, () = DEFAULT |
|---|---|---|
| Unique identifier watch | string | Example "D8F8" |
| Accelerometer interval | integer | 0 indicates switched off, 10–1000 (25) ms |
| Linear accelerometer interval | integer | 0 indicates switched off, 10–1000 (25) ms |
| Gyroscope interval | integer | 0 indicates switched off, 10–1000 (25) ms |
| Barometer interval | integer | 0 indicates switched off, 10–1000 (100) ms |
| GPS interval | integer | 1–10 (1) seconds |
| GPS middle point privacy circle, latitude | float | –90.0–90.0 (52.169311) degrees |
| GPS middle point privacy circle, longitude | float | –180.0–180.0 (4.456711) degrees |
| GPS radius privacy circle | integer | 0 indicates switched off, 10–1000 (100) meter |
| Write recordings interval | float | 0.01–10.0 (0.05) seconds |

**Table 2** Configuration settings.

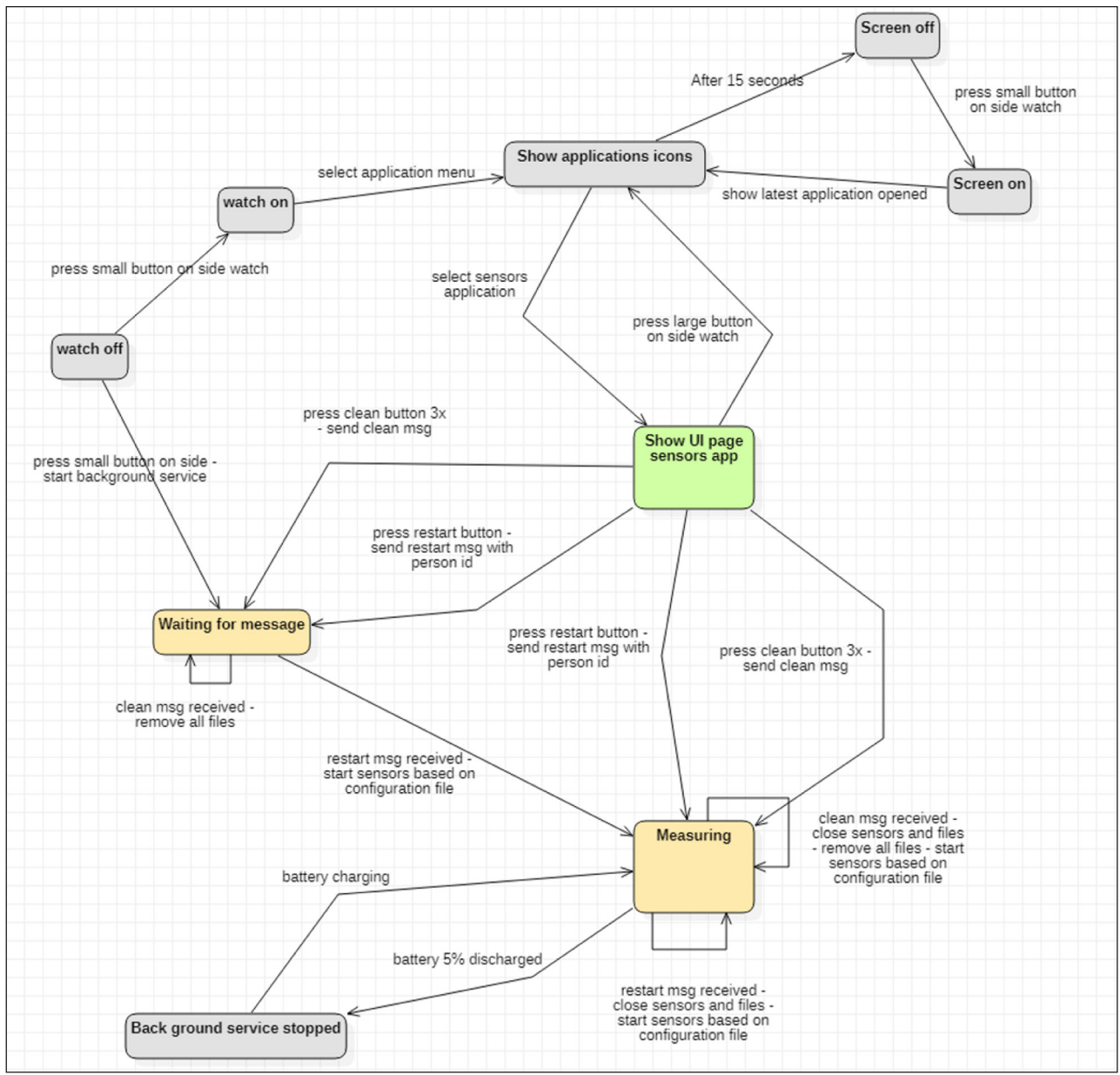

**Figure 2** State diagram of sensor application (green), sensor service (orange), and the Tizen OS (grey).

## 3 QUALITY CONTROL

The WEARDA software package and the sensor-value recordings of the Samsung smartwatch were manually tested because of the absence of a test framework and validated. The following paragraphs show validation of the accelerometer including gravity and linear acceleration and for the gravity acceleration the average value and standard deviation of several identical smartwatches; validation of the gyroscope, barometer, and GPS. Additionally, mitigation measures for the irregular sampling behavior of the watch are described.

### ACCELERATION AND GYROSCOPE
We conducted a spinning test — watch rotating on its Gorilla glass, decelerating in rotation speed – to test the main flow of WEARDA producing the expected raw data files following the configured settings, and validate the values of the (linear) acceleration and gyroscope sensors by inspecting the expected values along time, see Figures 3, 4 and 5. The sensors are positioned a bit outside of the center of the watch.

### GYROSCOPE AND ACCELEROMETER
The gyroscope values showed a decreasing trend on two of its axes, the gravity acceleration axis showed $ax = 0$, $ay = 1.2$ m/s$^2$, $az = 9.81$ m/s$^2$. The linear acceleration was not valid for high rotation velocities but showed promising values after 32 seconds. The linear acceleration is derived from the gravity acceleration with an algorithm used by Samsung.

### BASELINE MEASUREMENTS
Zero measurements or baseline measurements showed that the standard deviation of the gravitational acceleration $G = \sqrt{(ax^2 + ay^2 + az^2)}$ ranges between 0.02 – 0.036 m/$s^2$. Its average $G$ ranges between 9.6 – 10.2 m/$s^2$. Further, the average gravitational acceleration $G$ was dependent on the spatial orientation or spatial positioning of the watch compared to the direction of the gravitational force or orientation of the surface of the ground as well.

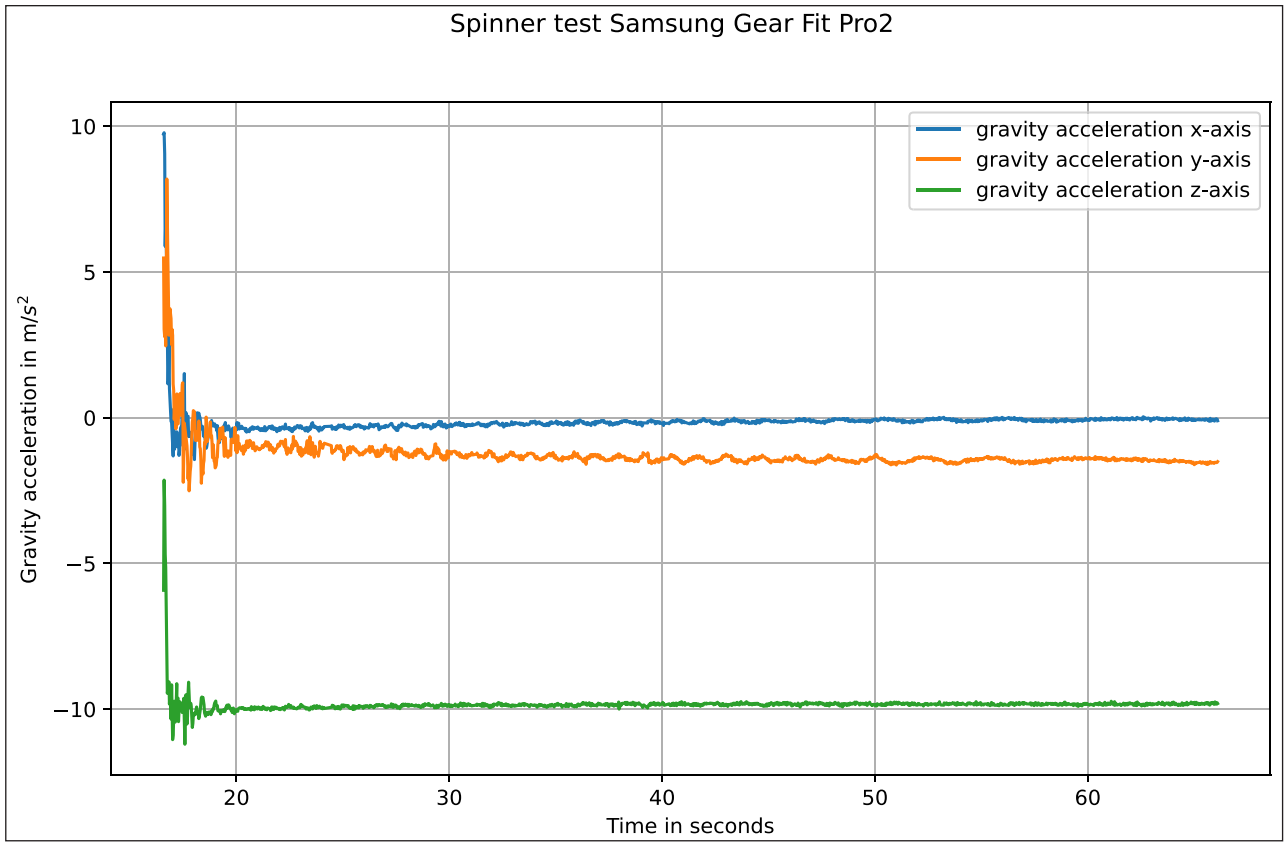

**Figure 3** Watch spinning on its glass, decelerating. The oscillation is caused by the movements of the arms of the band. The graph shows the gravity acceleration per axis in m/$s^2$.

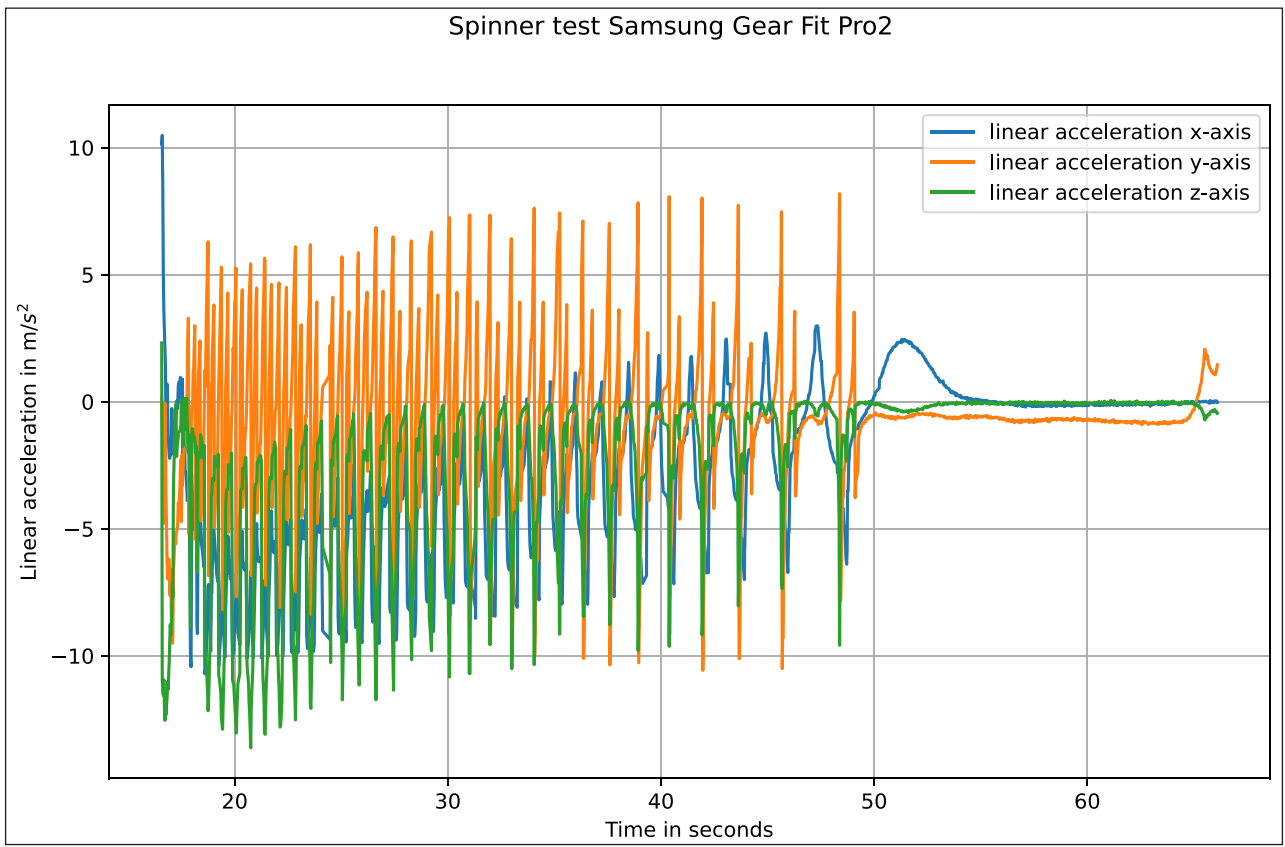

**Figure 4** Watch spinning on its glass, decelerating. The oscillation is caused by the movements of the arms of the band. The graph shows the linear acceleration per axis in m/$s^2$.

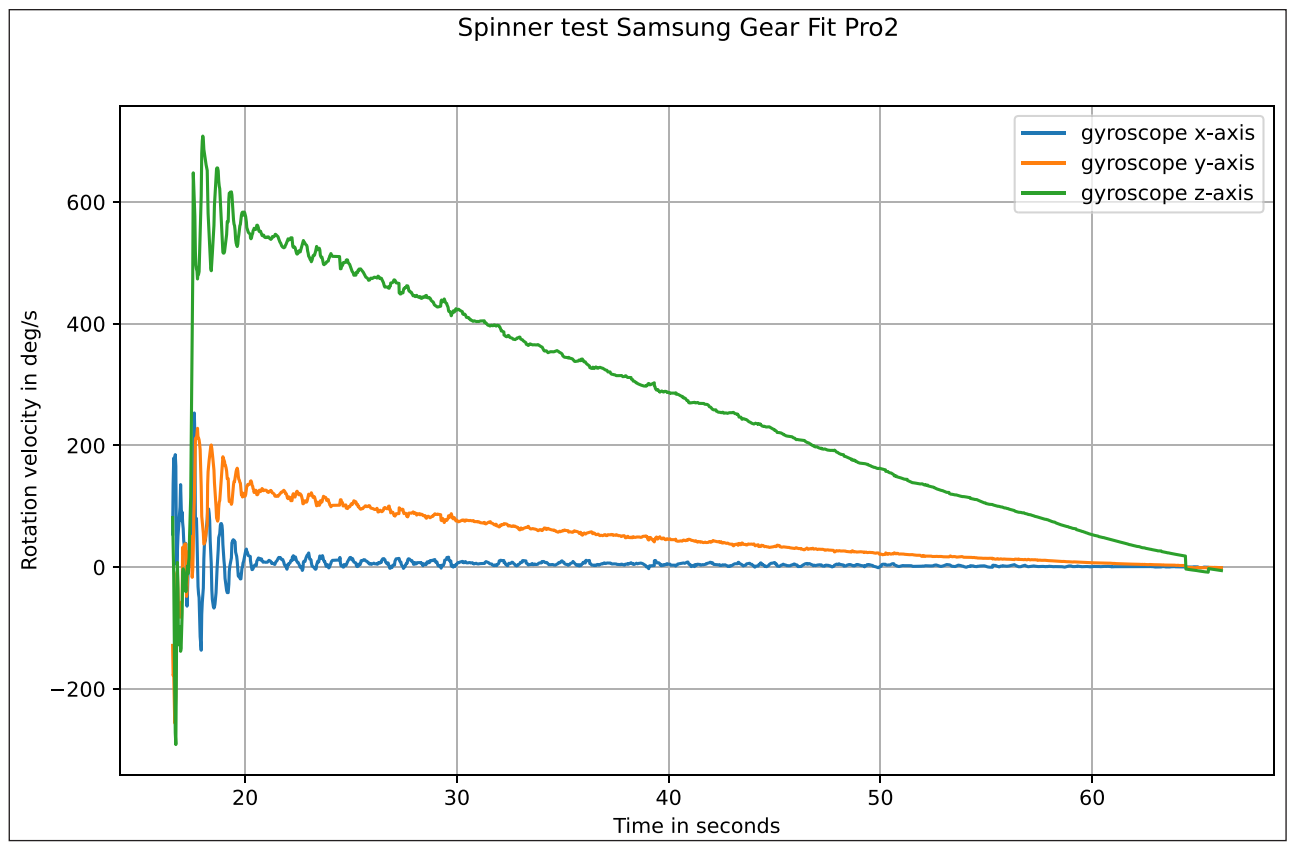

**Figure 5** Watch spinning on its glass, decelerating. The oscillation is caused by the movements of the arms of the band. The graph shows the rotation velocity per axis in degrees per second.

| SAMPLE RATE | SENSOR INTERVAL | WRITER INTERVAL | SAMPLE DENSITY GPS OFF | SAMPLE DENSITY GPS ON |
|---|---|---|---|---|
| 50 Hz | 10 ms | 20 ms | 0.99 | 0.90–0.93 |
| 50 Hz | 20 ms | 10 ms | 0.94 | |
| 40 Hz | 25 ms | 25 ms | 0.80–0.87 | 0.975 |
| 40 Hz | 10 ms | 25 ms | 0.86 | |
| 40 Hz | 15 ms | 25 ms | 0.94 | |
| 30 Hz | 15 ms | 33 ms | 0.96 | |
| 30 Hz | 33 ms | 33 ms | 0.97 | |
| 25 Hz | 20 ms | 40 ms | 0.99 | |
| 20 Hz | 10 ms | 50 ms | 0.96 | |
| 20 Hz | 25 ms | 50 ms | 0.98 | |
| 20 Hz | 50 ms | 50 ms | 0.93 | |
| 10 Hz | 100 ms | 100 ms | 0.99 | |

**Table 3** Sample density as a function of write- and sensor readout interval in milliseconds.

**BAROMETER AND GPS**
The barometer and GPS were validated by comparing the height differences of a GPS track with the air pressure measured.[5] An air pressure difference of 1 millibar (hPa) corresponded to 7.75 meters at sea level, temperature 10 degrees Celsius. The accuracy of the barometer is roughly 0.01 millibar (0.075 meters).

## DIFFERENT TIMESTAMP BIAS AND IRREGULAR SAMPLING RATE

We noticed that the Tizen OS API is producing different biased time stamps per sensor type. On top of that, the sampling time appears to be irregular.

We implemented a writing timer that writes all sensor values known at a certain moment, to one or more sensor files, to overcome these two problems. As a consequence, identical values can be written if the write timer interval is shorter than a sensor-specific readout interval. Therefore identical values are not written to the files. One could argue that a sensor value could be identical to its previous value. However, because of noise influences on the values, this probability is very low.

In case the sensor readout interval is shorter than the write timer interval, changing sensor values could be missed. We choose to write only the last available values.

## WRITER AND SENSOR INTERVALS, AND SAMPLE DENSITY

The best intervals of the writer timer and the sensor readouts are listed in Table 3. The table also contains the realized sample density which is defined as the ratio between recorded samples divided by the expected number of samples. A sample density of 1 is ideal; if it is lower than 1 the effective sample frequency is lower than the configured one; if it is larger than 1 the effective sample frequency is higher than the configured one.

# (2) AVAILABILITY

## OPERATING SYSTEM
Tizen OS developed by Samsung based on Linux OS, version 2.3.1:13

## PROGRAMMING LANGUAGE
C-programming language, Tizen API version 2.3.1:13

## ADDITIONAL SYSTEM REQUIREMENTS
The software needs some disk space to store the data collected. Typical for 7 days of data collection, 12 hours a day, 50 Hz sample frequency 500 MB will be sufficient.

## DEPENDENCIES
None

## LIST OF CONTRIBUTORS
Richard van Dijk, research software engineer of the LIACS Software Lab was the only developer of the software package and wrote most of the paper. The software was tested, refined functionally, and used in practice by Daniela Gawehns. Daniela Gawehns and Matthijs van Leeuwen contributed to the writing of this paper in a number of iterations and made contributions to the interpretation of the collected data for the work.

## SOFTWARE LOCATION
### Archive
**Name:** Zenodo
**Persistent identifier:** 10.5281/zenodo.7464332
**License:** MIT license
**Publisher:** LIACS Software Lab, Leiden University
**Version published:** v1.0.0
**Date published:** 2022/12/20

### Code repository
**Name:** GitHub
**Persistent identifier:** https://github.com/LiacsProjects/Wearda
**Licence:** MIT license
**Date published:** 2021/07/28

## LANGUAGE
The language of repository, documentation, software and supporting files, are all in English.

# (3) REUSE POTENTIAL

The WEARDA software package was initially created for the project "Dementia back at the heart of the community" [2].[6]

While developed for this project, the WEARDA package is designed in a generic manner to allow reuse in other contexts. Depending on the research, the configurations can be set to collect only the necessary data and at a frequency that is useful to the researcher (see Table 2 for an overview). The package can be used on other Tizen OS devices.

For projects where similar watches are used (i.e., smartwatches with a comparable set of sensors), the overall architecture of the WEARDA package is reusable, as are the optimization strategies for energy consumption, time synchronization, the robustness of the measurement, and storage of the data.

## AN EXAMPLE OF USE
WEARDA was successfully used to collect data at a nursing home in the Netherlands. Researchers wanted to assess how changes in care management affected residents' daily life and activity levels. The main focus was to find out how the new park surrounding the nursing home was used by residents. Therefore it was necessary to not only measure acceleration but also to record GPS trajectories to see if and how residents used the new park.

Researchers recorded activity (with accelerometers and gyroscope) and displacement (GPS traces) on five consecutive days. During the first two days, all participants were given watches configured to record sensor data as well as GPS data. On the remaining three days, only residents who have left the building during the first two days wore a watch with GPS recording. This reduced the need to switch wearables during lunchtime and recharge them as the continuous GPS recording needed a lot of battery power.

## DATA ACCESSIBILITY STATEMENT

The data collection of the Dementia project of LIACS and Nivel will be available only in highly aggregated form to other researchers but not to the general public because of their privacy-sensitive nature.

## NOTES

1. https://github.com/liacsprojects/wearda.
2. This project was a collaboration between the Netherlands Institute for Health Services Research (NIVEL), Utrecht and Leiden Institute of Advanced Computer Science (LIACS), Leiden University, funded by ZonMw, with project timeline from 2018 to 2022.
3. https://www.tizen.org.
4. https://www.samsung.com/us/business/support/owners/product/gear-fit2-pro-bluetooth.
5. https://www.gpsvisualizer.com.
6. This project was a collaboration between the Netherlands Institute for Health Services Research (NIVEL) Utrecht and Leiden Institute of Advanced Computer Science (LIACS), Leiden University, funded by ZONMW, with project timeline from 2018 to 2022.

## ACKNOWLEDGEMENTS

We acknowledge contributions from Renelle Bourdage, Jeremie Gobeil, and Joost Visser during the genesis of this project.

## FUNDING INFORMATION

This work is partly financed by ZonMw, under project number 733050846, the hours of the LIACS Software Lab were financed by LIACS, the Leiden Institute of Advanced Computer Science.

## COMPETING INTERESTS

The authors have no competing interests to declare.

## AUTHOR AFFILIATIONS

**Richard M.K. van Dijk**  orcid.org/0000-0002-5796-7883
Research software engineer, Leiden Institute of Advanced Computer Science, Faculty of Science, Leiden University, The Netherlands

**Daniela Gawehns**  orcid.org/0000-0002-9678-9012
PhD candidate, Leiden Institute of Advanced Computer Science, Faculty of Science, Leiden University, The Netherlands

**Matthijs van Leeuwen**  orcid.org/0000-0002-0510-3549
Associate professor, Leiden Institute of Advanced Computer Science, Faculty of Science, Leiden University, The Netherlands

## REFERENCES

1. **Breasail MO,** et al. "Wearable GPS and Accelerometer Technologies for Monitoring Mobility and Physical Activity in Neurodegenerative Disorders: A Systematic Review". en. In: *Sensors* 21.24 (Jan. 2021); Number: 24 Publisher: Multidisciplinary Digital Publishing Institute, p. 8261. URL: https://www.mdpi.com/1424-8220/21/24/8261 (visited on 01/07/2022). DOI: https://doi.org/10.3390/s21248261

2. **"Dementia back at the heart of the community".** In: Website Leiden University (). URL: https://www.universiteitleiden.nl/en/research/research-projects/data-science-research-programme/data-science-research-programme-project-daniela-gawehns.

3. **Huhn S,** et al. "The Impact of Wearable Technologies in Health Research: Scoping Review". en. In: *JMIR mHealth and uHealth* 10.1 (Jan. 2022); e34384. ISSN: 2291-5222. URL: https://mhealth.jmir.org/2022/1/e34384 (visited on 11/30/2022). DOI: https://doi.org/10.2196/34384

4. **Menai M,** et al. "Accelerometer assessed moderate-to-vigorous physical activity and successful ageing: results from the Whitehall II study". en. In: *Scientific Reports* 7.1 (May 2017); p. 45772. ISSN: 2045-2322. URL: http://www.nature.com/articles/srep45772 (visited on 01/07/2022). DOI: https://doi.org/10.1038/srep45772

5. **Millar GC,** et al. "Space-time analytics of human physiology for urban planning". en. In: *Computers, Environment and Urban Systems* 85 (Jan. 2021); p. 101554. ISSN: 0198-9715. URL: https://www.sciencedirect.com/science/article/pii/S0198971520302878 (visited on 11/30/2022). DOI: https://doi.org/10.1016/j.compenvurbsys.2020.101554

6. **Suri A,** et al. "Facilitators and barriers to real-life mobility in community-dwelling older adults: a narrative review of accelerometry- and global positioning system-based studies". en. In: *Aging Clinical and Experimental Research* 34.8 (Aug. 2022); Company: Springer Distributor: Springer Institution: Springer Label: Springer Number: 8 Publisher: Springer International Publishing, pp. 1733–1746. ISSN: 1720-8319. URL: http://link.springer.com/article/10.1007/s40520-022-02096-x (visited on 11/30/2022). DOI: https://doi.org/10.1007/s40520-022-02096-x

7. **Tribby CP,** et al. "Assessing built environment walkability using activity-space summary measures". en. In: *Journal of Transport and Land Use* (June 2015); ISSN: 1938-7849. URL: https://www.jtlu.org/index.php/jtlu/article/view/625

(visited on 11/30/2022). DOI: https://doi.org/10.5198/jtlu.2015.625

8. **Welch V,** et al. "Use of Mobile and Wearable Artificial Intelligence in Child and Adolescent Psychiatry: Scoping Review". en. In: *Journal of Medical Internet Research* 24.3 (Mar. 2022); e33560. ISSN: 1438-8871. URL: https://www.jmir.org/2022/3/e33560 (visited on 11/30/2022). DOI: https://doi.org/10.2196/33560

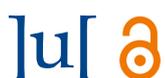